\documentclass[12pt]{article}
\usepackage{amssymb}

\input{tcilatex}

\begin{document}

\bigskip\ 

\bigskip\ 

\bigskip\ 

\begin{center}
\textbf{DUALITY SYMMETRY IN KALUZA-KLEIN}

$n+D+d$ \textbf{DIMENSIONAL COSMOLOGICAL MODEL}

\textbf{\ }

\smallskip\ 

J. A. Nieto$^{a}$\footnote{%
nieto@uas.uasnet.mx}M. P. Ryan$^{b}$, O. Velarde$^{a}$ and C. M. Yee$^{a}$

\bigskip\ 

$^{a}$\textit{Facultad de Ciencias F\'{\i}sico-Matem\'{a}ticas de la
Universidad Aut\'{o}noma}

\textit{de Sinaloa, C.P. 80010, Culiac\'{a}n Sinaloa, M\'{e}xico}

\smallskip\ 

$^{b}$\textit{Instituto de Ciencias Nucleares, UNAM, A. Postal 70-543,}

\textit{M\'{e}xico 04510 D. F. M\'{e}xico}

\bigskip\ 

\textbf{Abstract}
\end{center}

It is shown that, with the only exception of $n=2,$ the Einstein-Hilbert
action in $n+D+d$ dimensions, with $n$ times, is invariant under the duality
transformation $a\rightarrow \frac{1}{a}$ and $b\rightarrow \frac{1}{b}$,
where $a$ is a Friedmann-Robertson-Walker scale factor in $D$ dimensions and 
$b$ a Brans-Dicke scalar field in $d$ dimensions respectively. We
investigate the $2+D+d$ dimensional cosmological model in some detail.

\bigskip\ 

\bigskip\ 

\bigskip\ 

\bigskip\ 

\bigskip\ 

Key words; Cosmological models, Kaluza-Klein theory, two time physics.

Pacs No.: 98.80.Dr, 12.10.Gq, 04.50.+h.

February 2004

\newpage

\noindent \textbf{1.-INTRODUCTION}

\smallskip\ 

It is known that string theory [1] suggests for the curvature a dual
behavior around time axis [2]. From a cosmological point of view, this
important property of string theory appears as a duality symmetry of the
effective action associated to a cosmological model . At present, string
theory is, however, thought as a limit of a bigger theory called M-theory
[3]. Therefore, it may be interesting to see if the cosmological duality
symmetry for string theory is also a symmetry for the M-theory.

There are two main proposals for M-theory: Matrix-theory [4] and N=2 string
theory [5]. These two scenarios may, in fact, be related [6]. In particular
N=2 string theory requires twelve dimensions with two time physics [7]. In
fact, two time physics have a number of interesting physical features.
First, all possible gravitational theories and gauge theories can be
understood from the perspective of two time physics [8]. Second, duality
symmetries in M-theory can be understood as a gauge symplectic
transformations. Finally, a part from M-theory, two time physics is
interesting for its own right because it may allow to discover new
symmetries in ordinary one time gauge theories. This scenario has also been
considered from the point of view of dualities involving compactifications
on timelike circles as well as spacelike circle ones [9]. In particular, it
has been shown that T-duality on a timelike circle takes type IIA theory
into a type IIB$^{\ast }$ theory and type IIB$^{\ast }$ theory into a type
IIA theory and that the strong-coupling limit of type IIA$^{\ast }$ is a
theory in 9+2 dimensional theory, denoted by M$^{\ast }$.

In general, we may expect that the theory of everything predicts not only
the dimensionality of the spacetime but also its signature. Motivated by
this observation, some authors [10] have considered a unify model in which
the world has $n$ arbitrary time-like dimensions and $s$ arbitrary
space-like dimensions.

Taking $n$ time physics seriously we may assume the Einstein-Hilbert action
in $n+s$ dimensions. Of course, for a realistic model we should compactify $%
d $ dimensions out of the $s$ space-like dimensions. In this scenario, the
easiest cosmological model is a Friedmann-Robertson-Walker-Brans-Dicke model
with scale factor $a$ in $D=s-d$ dimensions and scale factor $b$ in $d$
dimensions. In this work, following the Tkach \textit{et al} formalism [11]
we investigate the duality symmetry $a\rightarrow \frac{1}{a}$ and $%
b\rightarrow \frac{1}{b}$ in such a cosmological model. We find that the
Einstein-Hilbert action in $n+D+d$ dimensions is invariant under such a
duality symmetry except in the case $n=2$. In general, an exceptional result
is very attractive in unified field theories since it selects a physical
model out of many other possibilities. For this reason we investigate the $%
2+D+d$ cosmological model in some detail.

The plan of this work is as follows. In section 2, we briefly review a
cosmological model in $1+D+d$ dimensions and in section 3, we apply such a
model to the $1+3+1$ dimensional theory. In section 4, we show that a
Friedmann-Robertson-Walker-Brans-Dicke cosmological model in $n+D+d$
dimensions is invariant under the duality symmetry $a\rightarrow \frac{1}{a}$
and $b\rightarrow \frac{1}{b}$, with the only exception of $n=2$. In section
5, we study in some detail a $2+D+d$ dimensional cosmological model.
Finally, in section 6, we make some final remarks.

\bigskip\ 

\noindent \textbf{2.-\ A }$1+D+d$\textbf{\ COSMOLOGICAL MODEL}

\smallskip\ 

Consider a universe described by the line element

\begin{equation}
ds^{2}=-N^{2}(t)dt^{2}+a^{2}(t)d^{D}\Omega +b^{2}(t)d^{d}\Sigma ,  \tag{1}
\end{equation}%
where $d^{D}\Omega $ and $d^{d}\Sigma $ correspond to a $D$-dimensional and $%
d$-dimensional homogenous spatial spaces with constant curvature $%
k_{1}=0,\pm 1$ and $k_{2}=0,\pm 1$ respectively.

Using (1) we find that the Einstein-Hilbert action in $\mathcal{D}=1+s$
dimensions, with $s=D+d$,

\begin{equation}
S=-\frac{1}{V_{s}}\int d^{\mathcal{D}}x\sqrt{-g}(R-2\Lambda ),  \tag{2}
\end{equation}%
where $V_{s}$ is an appropriate volume constant, becomes (see Appendix A)

\begin{equation}
\begin{array}{c}
S=-\int dtNa^{D}b^{d}\{2DN^{-2}a^{-1}\ddot{a}-2DN^{-3}\dot{N}a^{-1}\dot{a}%
+D(D-1)N^{-2}a^{-2}\dot{a}^{2} \\ 
\\ 
+D(D-1)k_{1}a^{-2}+2dN^{-2}b^{-1}\ddot{b}-2dN^{-3}\dot{N}b^{-1}\dot{b}%
+d(d-1)N^{-2}b^{-2}\dot{b}^{2} \\ 
\\ 
+d(d-1)k_{2}b^{-2}+2dDN^{-2}a^{-1}\dot{a}b^{-1}\dot{b}-2\Lambda \}.%
\end{array}
\tag{3}
\end{equation}%
Here, we have performed a volume integration over the space-like
coordinates. The action (3) can be rewritten as

\begin{equation}
\begin{array}{c}
S=-\int dt\{[\frac{d}{dt}(2DN^{-1}a^{D-1}b^{d}\dot{a}+2dN^{-1}a^{D}b^{d-1}%
\dot{b})-D(D-1)N^{-1}a^{D-2}b^{d}\dot{a}^{2} \\ 
\\ 
-d(d-1)N^{-1}a^{D}b^{d-2}\dot{b}^{2}-2dDN^{-1}a^{D-1}\dot{a}b^{d-1}\dot{b}]
\\ 
\\ 
+D(D-1)k_{1}Na^{D-2}b^{d}+d(d-1)k_{2}Na^{D}b^{d-2}-2\Lambda Na^{D}b^{d}\}.%
\end{array}
\tag{4}
\end{equation}%
Since a total derivative does not contribute to the dynamics of the
classical system we can drop the first term in (4). Thus (4) simplifies to

\begin{equation}
\begin{array}{c}
S=\int dt\{N^{-1}a^{D}b^{d}[D(D-1)a^{-2}\dot{a}^{2}+d(d-1)b^{-2}\dot{b}%
^{2}+2dDa^{-1}\dot{a}b^{-1}\dot{b}] \\ 
\\ 
-D(D-1)k_{1}Na^{D-2}b^{d}-d(d-1)k_{2}Na^{D}b^{d-2}+2\Lambda Na^{D}b^{d}\}.%
\end{array}
\tag{5}
\end{equation}

Consider now the duality transformation

\begin{equation}
\begin{array}{ccc}
a & \rightarrow & \frac{1}{a}, \\ 
&  &  \\ 
b & \rightarrow & \frac{1}{b}, \\ 
&  &  \\ 
N & \rightarrow & Na^{-2D}b^{-2d}.%
\end{array}
\tag{6}
\end{equation}%
We find that under this transformation the action (5) becomes

\begin{equation}
\begin{array}{ccc}
S & \rightarrow & S=\int dt\{N^{-1}a^{D}b^{d}[D(D-1)a^{-2}\dot{a}%
^{2}+d(d-1)b^{-2}\dot{b}^{2} \\ 
&  &  \\ 
&  & +2dDa^{-1}\dot{a}b^{-1}\dot{b}]-D(D-1)k_{1}Na^{-3D+2}b^{-3d} \\ 
&  &  \\ 
&  & -d(d-1)k_{2}Na^{-3D}b^{-3d+2}+2\Lambda Na^{-3D}b^{-3d}\}.%
\end{array}
\tag{7}
\end{equation}%
We learn from (7) that only if the constants $k_{1},k_{2}$ and $\Lambda $
are zero the action (7) remains invariant under the duality transformation
(6).

\bigskip\ 

\noindent \textbf{3.-} \textbf{A }$1+3+1$\textbf{\ DIMENSIONAL COSMOLOGICAL
MODEL}

\smallskip\ 

We now consider a particular case of the model described in the previous
section. We assume a universe described by a homogeneous and isotropic
Friedmann-Robertson-Walker metric in five dimensions

\begin{equation}
ds^{2}=-N^{2}(t)dt^{2}+a^{2}(t)d^{3}\Omega +b^{2}(t)dx^{4},  \tag{8}
\end{equation}%
where $d^{3}\Omega $ is the interval on the spatial sector with constant
curvature $k=0,\mp 1,$ corresponding to plane, hyperbolic or spherical
three-space, respectively.

In the case of $\mathcal{D}=1+(3+1)$ dimensions the Einstein-Hilbert action
(2) becomes

\begin{equation}
S=-\frac{1}{6V_{5}}\int d^{5}x\sqrt{-g}(R-2\Lambda ),  \tag{9}
\end{equation}%
where $V_{5}$ is an appropriate volume constant. Considering (8) we observe
that, after performing a volume integration over the space-like coordinates,
the action (9) leads to (see Appendix B)%
\begin{equation}
\begin{array}{c}
S=-\frac{1}{6}\int dtNa^{3}b\{6N^{-2}a^{-1}\ddot{a}-6N^{-3}\dot{N}a^{-1}\dot{%
a}+6N^{-2}a^{-2}\dot{a}^{2} \\ 
\\ 
+6ka^{-2}+2N^{-2}b^{-1}\ddot{b}-2N^{-3}\dot{N}b^{-1}\dot{b}+6N^{-2}\dot{N}%
a^{-1}\dot{a}b^{-1}\dot{b}-2\Lambda \}.%
\end{array}
\tag{10}
\end{equation}%
This action can be rewritten as

\begin{equation}
\begin{array}{c}
S=-\frac{1}{6}\int dt\{6[\frac{d}{dt}(N^{-1}a^{2}\dot{a}b)-N^{-1}a\dot{a}%
^{2}b-N^{-1}a^{2}\dot{a}\dot{b}] \\ 
\\ 
+2[\frac{d}{dt}(N^{-1}a^{3}\dot{b})]+6kNab-2\Lambda Na^{3}b\}.%
\end{array}
\tag{11}
\end{equation}%
Dropping the total derivatives from (11), we see that this expression
simplifies to

\begin{equation}
\begin{array}{c}
S=\int dt\{N^{-1}a^{3}b[a^{-2}\dot{a}^{2}+a^{-1}\dot{a}b^{-1}\dot{b}]-kNab+%
\frac{1}{3}\Lambda Na^{3}b\}.%
\end{array}
\tag{12}
\end{equation}

Consider now the duality transformation

\begin{equation}
\begin{array}{ccc}
a & \rightarrow & \frac{1}{a}, \\ 
&  &  \\ 
b & \rightarrow & \frac{1}{b}, \\ 
&  &  \\ 
N & \rightarrow & Na^{-6}b^{-2}.%
\end{array}
\tag{13}
\end{equation}%
We find that under this transformation the action (12) transforms as

\begin{equation}
\begin{array}{ccc}
S & \rightarrow & S=\int dt\{N^{-1}a^{3}b[a^{-2}\dot{a}^{2}+a^{-1}\dot{a}%
b^{-1}\dot{b}]-kNa^{-7}b^{-3}+\frac{1}{3}\Lambda Na^{-9}b^{-3}\}.%
\end{array}
\tag{14}
\end{equation}%
We again observe that only if $k$ and $\Lambda $ are zero the action (14)
remains invariant under the duality transformation (13).

The equations of motion derived from (14) are%
\begin{equation}
\begin{array}{c}
b^{-1}\ddot{b}+2a^{-1}\ddot{a}+2a^{-1}\dot{a}b^{-1}\dot{b}+a^{-2}\dot{a}%
^{2}-N^{-1}\dot{N}(2a^{-1}\dot{a}+b^{-1}\dot{b}) \\ 
\\ 
+kN^{2}a^{-2}-\Lambda N^{2}=0, \\ 
\\ 
a^{-1}\ddot{a}+a^{-2}\dot{a}^{2}-N^{-1}\dot{N}a^{-1}\dot{a}+kN^{2}a^{-2}-%
\frac{1}{3}\Lambda N^{2}=0, \\ 
\\ 
a^{-1}\dot{a}b^{-1}\dot{b}+a^{-2}\dot{a}^{2}+kN^{2}a^{-2}-\frac{1}{3}\Lambda
N^{2}=0.%
\end{array}
\tag{15}
\end{equation}%
If we now redefine the time as $d\tau =Ndt$ and define $a^{^{\prime }}=\frac{%
da}{d\tau }$ we have that the equations (15) simplify to

\begin{equation}
\begin{array}{c}
b^{-1}b^{^{\prime \prime }}+2a^{-1}a^{^{\prime \prime }}+2a^{-1}a^{^{\prime
}}b^{-1}b^{^{\prime }}+a^{-2}a^{^{\prime }2}+ka^{-2}-\Lambda =0, \\ 
\\ 
a^{-1}a^{^{\prime \prime }}+a^{-2}a^{^{\prime }2}+ka^{-2}-\frac{1}{3}\Lambda
=0, \\ 
\\ 
a^{-1}a^{^{\prime }}b^{-1}b^{^{\prime }}+a^{-2}a^{^{\prime }2}+ka^{-2}-\frac{%
1}{3}\Lambda =0.%
\end{array}
\tag{16}
\end{equation}%
It is not difficult to show that from the last two equations in (16) one
obtains the formula

\begin{equation}
a^{^{\prime }}=\beta b,  \tag{17}
\end{equation}%
where $\beta $ is an integration constant.

If $\Lambda =0$ we get the solution [12];

\begin{equation}
a(\tau )=k^{\frac{1}{2}}\sqrt{a_{0}^{2}-(\tau -a_{0})^{2}},  \tag{18}
\end{equation}
where $a_{0}$ is the value of $a$ at the classical turning point. We also
have

\begin{equation}
b(\tau )=\frac{-k^{\frac{1}{2}}(\tau -a_{0})}{\beta \sqrt{a_{0}^{2}-(\tau
-a_{0})^{2}}}.  \tag{19}
\end{equation}%
While if $\Lambda \neq 0$ we have the solution

\begin{equation}
a^{2}(\tau )=C_{1}\exp (\alpha (\frac{2}{3}\mid \Lambda \mid )^{\frac{1}{2}%
}\tau ))+C_{2}\exp (-\alpha (\frac{2}{3}\mid \Lambda \mid )^{\frac{1}{2}%
}\tau ))+\frac{3k}{\Lambda },  \tag{20}
\end{equation}%
where $C_{1}$ and $C_{2}$ are arbitrary constants and $\alpha $ is equal to $%
i$ or $1$ depending if $\Lambda <0$ or $\Lambda >0.$ We also have

\begin{equation}
b(\tau )=\frac{\alpha (\frac{2}{3}\mid \Lambda \mid )^{\frac{1}{2}%
}\{C_{1}\exp (\alpha (\frac{2}{3}\mid \Lambda \mid )^{\frac{1}{2}}\tau
))-C_{2}\exp (-\alpha (\frac{2}{3}\mid \Lambda \mid )^{\frac{1}{2}}\tau ))\}%
}{2\beta \sqrt{C_{1}\exp (\alpha (\frac{2}{3}\mid \Lambda \mid )^{\frac{1}{2}%
}\tau ))+C_{2}\exp (-\alpha (\frac{2}{3}\mid \Lambda \mid )^{\frac{1}{2}%
}\tau ))+\frac{3k}{\Lambda }}}.  \tag{21}
\end{equation}%
Notice that these solutions are written in terms of the parameter $\tau $
and not in terms of the original time $t$. This is not a real problem
because using the transformation $d\tau =Ndt$ we can always go from $\tau $
to $t$ and vice versa. Alternatively, we can understand the parameter $\tau $
as the parameter $t$ with $N=1$ which it means that we have fixed the time
gauge parameter.

Let us now discuss the quantum theory of this model. From (12) we have the
lagrangian

\begin{equation}
L=N^{-1}a^{3}b[a^{-2}\dot{a}^{2}+a^{-1}\dot{a}b^{-1}\dot{b}]-kNab+\frac{1}{3}%
\Lambda Na^{3}b.  \tag{22}
\end{equation}%
From this lagrangian we get that the linear momenta associated to $a$ and $b$
are

\begin{equation}
P_{a}=\frac{\partial L}{\partial \dot{a}}=N^{-1}a(2b\dot{a}+a\dot{b}) 
\tag{23}
\end{equation}
and

\begin{equation}
P_{b}=\frac{\partial L}{\partial \dot{b}}=N^{-1}a^{2}\dot{a},  \tag{24}
\end{equation}
respectively.

Using (22)-(24) we obtain that the hamiltonian $H=\dot{a}P_{a}+\dot{b}%
P_{b}-L $ becomes

\begin{equation}
H=N[a^{-2}(P_{a}P_{b}-a^{-1}bP_{b}^{2})+kab-\frac{1}{3}\Lambda a^{3}b]. 
\tag{25}
\end{equation}
Since under (13) the linear momenta $P_{a}$ and $P_{b}$ transform as

\begin{equation}
\begin{array}{ccc}
P_{a} & \rightarrow & -\frac{1}{a^{-2}}P_{a}, \\ 
&  &  \\ 
P_{b} & \rightarrow & -\frac{1}{b^{-2}}P_{b},%
\end{array}
\tag{26}
\end{equation}%
we note that only if $k$ and $\Lambda $ are zero the hamiltonian (25) is
invariant under the duality transformations (13) and (26).

Thus, considering (22) and (25) we learn that the first order lagrangian is

\begin{equation}
L=\dot{a}P_{a}+\dot{b}P_{b}-N[a^{-2}(P_{a}P_{b}-a^{-1}bP_{b}^{2})+kab-\frac{1%
}{3}\Lambda a^{3}b].  \tag{27}
\end{equation}%
Note that $N$ acts as a lagrange multiplier. Therefore, we have the
constraint

\begin{equation}
a^{-2}(P_{a}P_{b}-a^{-1}bP_{b}^{2})+kab-\frac{1}{3}\Lambda a^{3}b=0, 
\tag{28}
\end{equation}%
which, according to the Dirac's constraint hamiltonian formalism,
annihilates the physical states $\mid \psi >$ at the quantum level

\begin{equation}
\lbrack a^{-2}(\hat{P}_{a}\hat{P}_{b}-a^{-1}b\hat{P}_{b}^{2})+kab-\frac{1}{3}%
\Lambda a^{3}b]\mid \psi >=0,  \tag{29}
\end{equation}%
where we promoted the linear momenta $P_{a}$ and $P_{b}$ as the operators $%
\hat{P}_{a}$ and $\hat{P}_{b}$, respectively. In the coordinate
representation we can choose $\hat{P}_{a}=-i\frac{\partial }{\partial a}$
and $\hat{P}_{b}=-i\frac{\partial }{\partial b}.$

For the particular case in which $k=\Lambda =0$ the Wheeler-DeWitt's
equation derived from (29) is%
\begin{equation}
-a\frac{{\partial ^{2}\psi }}{{\partial }a{\partial }b}+b\frac{{\partial
^{2}\psi }}{{\partial }b{^{2}}}=0,  \tag{30}
\end{equation}%
where we assumed the simplest normal ordering. This equation is separable
and therefore we can propose a solution of the form

\begin{equation}
\psi =A(a)B(b).  \tag{31}
\end{equation}%
We have%
\begin{equation}
-\frac{a}{A}\frac{dA}{da}+\frac{b}{\frac{dB}{db}}\frac{d^{2}B}{db^{2}}=0 
\tag{32}
\end{equation}%
or%
\begin{equation}
b\frac{1}{\frac{dB}{db}}\frac{d^{2}B}{db^{2}}=-q=\frac{a}{A}\frac{dA}{da}. 
\tag{33}
\end{equation}%
where $q$ is a new variable independent of $a$ and $b$. The general solution
for $\psi $ is%
\begin{equation}
\psi =\int [C_{1}(q)be^{-q\ln (ab)}+C_{2}(q)e^{-q\ln a}]dq.  \tag{34}
\end{equation}%
Since $C_{1}$ and $C_{2}$ are arbitrary functions we find that the general
solution for $\psi $ can be rewritten in the form%
\begin{equation}
\psi =bF_{1}(ab)+F_{2}(a),  \tag{35}
\end{equation}%
where $F_{1}$ and $F_{2}$ are arbitrary functions.

\bigskip\ 

\bigskip\ 

\bigskip\ 

\noindent \textbf{4.-} \textbf{A }$n+D+d$\textbf{\ DIMENSIONAL COSMOLOGICAL
MODEL}

\smallskip\ 

Consider the line element

\begin{equation}
ds^{2}=g_{AB}(x^{C})dx^{A}dx^{B}+a^{2}(x^{C})d^{D}\Omega
+b^{2}(x^{C})d^{d}\Sigma ,  \tag{36}
\end{equation}%
where the indices $A,B$ run from $1$ to $n$.

For the line element (36), we have that the action

\begin{equation}
S=-\frac{1}{V_{D+d}}\int d^{n+D+d}x\sqrt{-g}(R-2\Lambda )  \tag{37}
\end{equation}%
leads to (see Appendix C)

\begin{equation}
\begin{array}{c}
S=-\int d^{n}x\sqrt{\bar{g}}a^{D}b^{d}\{-2Da^{-1}D_{A}\partial
^{A}a-D(D-1)g^{AB}a^{-2}\partial _{A}a\partial _{B}a \\ 
\\ 
-2db^{-1}D_{A}\partial ^{A}b-d(d-1)g^{AB}b^{-2}\partial _{A}b\partial _{B}b
\\ 
\\ 
-2Dd(b^{-1}a^{-1})g^{AB}\partial _{A}a\partial _{B}b+\bar{R}+a^{-2}\tilde{R}%
+b^{-2}\hat{R}-2\Lambda \}.%
\end{array}
\tag{38}
\end{equation}%
This action can be rewritten as

\begin{equation}
\begin{array}{c}
S=-\int d^{n}x\sqrt{\bar{g}}\{D_{A}(2Da^{D-1}\partial
^{A}ab^{d}+2db^{d-1}\partial ^{A}ba^{D}) \\ 
\\ 
-D(D-1)g^{AB}a^{D-2}b^{d}\partial _{A}a\partial
_{B}a-d(d-1)g^{AB}a^{D}b^{d-2}\partial _{A}b\partial _{B}b \\ 
\\ 
-2Dd(a^{D-1}b^{d-1})g^{AB}\partial _{A}a\partial _{B}b+a^{D}b^{d}(\bar{R}%
+a^{-2}\tilde{R}+b^{-2}\hat{R}-2\Lambda )\}.%
\end{array}
\tag{39}
\end{equation}%
Therefore, we obtain

\begin{equation}
\begin{array}{c}
\int d^{n}x\sqrt{\bar{g}}a^{D}b^{d}\{D(D-1)g^{AB}a^{-2}\partial
_{A}a\partial _{B}a+d(d-1)g^{AB}b^{-2}\partial _{A}b\partial _{B}b \\ 
\\ 
+2Dd(a^{-1}b^{-1})g^{AB}\partial _{A}a\partial _{B}b-(\bar{R}+a^{-2}\tilde{R}%
+b^{-2}\hat{R}-2\Lambda )\}.%
\end{array}
\tag{40}
\end{equation}

Now, it is straightforward to verify that (40) is invariant under the
duality transformation

\begin{equation}
\begin{array}{ccc}
a & \rightarrow & \frac{1}{a}, \\ 
&  &  \\ 
b & \rightarrow & \frac{1}{b}, \\ 
&  &  \\ 
g_{AB} & \rightarrow & a^{\frac{4D}{n-2}}b^{\frac{4d}{n-2}}g_{AB}.%
\end{array}
\tag{41}
\end{equation}%
provided that $n\neq 2$ and $\bar{R}+a^{-2}\tilde{R}+b^{-2}\hat{R}-2\Lambda
=0.$ Observe that when $n=1$ the duality transformation (41) is reduced to
the particular case (6).

What appears interesting from our analysis of the invariance of (40) under
the duality transformation (41) is that the case $n=2$ is distinguished
among any other $n$ value. In other words, from duality point of view two
time physics turns out to be a singular case. In some sense duality symmetry
is playing analogue role in several time cosmological physics as the Weyl
invariance in p-brane physics (see [13] and references there in).

\bigskip\ 

\noindent \textbf{5.-} \textbf{A} \textbf{TWO TIME COSMOLOGICAL MODEL}

\smallskip\ 

The analysis of previous section may motivate us to study in some detail the
case of $n=2$. Let us apply (40) to the case of two time physics. In this
case, considering the change of variables

\[
\begin{array}{c}
a=e^{\lambda }, \\ 
\\ 
b=e^{\sigma },%
\end{array}%
\]%
we find that the action (40) becomes:

\begin{equation}
\begin{array}{c}
S=\int d^{2}x\sqrt{\bar{g}}e^{D\lambda }e^{d\sigma }\{D(D-1)g^{AB}\partial
_{A}\lambda \partial _{B}\lambda \\ 
\\ 
+d(d-1)g^{AB}\partial _{A}\sigma \partial _{B}\sigma +Ddg^{AB}\{\partial
_{A}\lambda \partial _{B}\sigma +\partial _{A}\sigma \partial _{B}\lambda \}
\\ 
\\ 
-(\bar{R}+e^{-2\lambda }\tilde{R}+e^{-2\sigma }\hat{R}-2\Lambda )\}.%
\end{array}
\tag{42}
\end{equation}%
Varying (42) with respect to $\lambda $ we get

\begin{equation}
\begin{array}{c}
\partial _{A}\{\sqrt{\bar{g}}g^{AB}e^{D\lambda }e^{d\sigma }[2D(D-1)\partial
_{B}\lambda +2Dd\partial _{B}\sigma ]\} \\ 
\\ 
-\sqrt{\bar{g}}e^{D\lambda }e^{d\sigma }\{D^{2}(D-1)g^{AB}\partial
_{A}\lambda \partial _{B}\lambda +Dd(d-1)g^{AB}\partial _{A}\sigma \partial
_{B}\sigma \\ 
\\ 
+D^{2}dg^{AB}[\partial _{A}\lambda \partial _{B}\sigma +\partial _{A}\sigma
\partial _{B}\lambda ] \\ 
\\ 
-[D\bar{R}+\left( D-2\right) e^{-2\lambda }\tilde{R}+De^{-2\sigma }\hat{R}%
-2D\Lambda ]\}=0,%
\end{array}
\tag{43a}
\end{equation}%
while varying $\sigma $ we obtain

\begin{equation}
\begin{array}{c}
\partial _{A}\{\sqrt{\bar{g}}g^{AB}e^{D\lambda }e^{d\sigma }[2d(d-1)\partial
_{B}\sigma +2Dd\partial _{B}\lambda ]\} \\ 
\\ 
-\sqrt{\bar{g}}e^{D\lambda }e^{d\sigma }\{d^{2}(d-1)g^{AB}\partial
_{A}\sigma \partial _{B}\sigma +Dd(D-1)g^{AB}\partial _{A}\lambda \partial
_{B}\lambda \\ 
\\ 
+Dd^{2}g^{AB}[\partial _{A}\lambda \partial _{B}\sigma +\partial _{A}\sigma
\partial _{B}\lambda ] \\ 
\\ 
-[d\bar{R}+de^{-2\lambda }\tilde{R}+\left( d-2\right) e^{-2\sigma }\hat{R}%
-2d\Lambda ]\}=0.%
\end{array}
\tag{43b}
\end{equation}%
Finally, varying (42) with respect to $g^{AB}$ we have

\begin{equation}
\begin{array}{c}
D\left( D-1\right) [\partial _{A}\lambda \partial _{B}\lambda -\frac{1}{2}%
g_{AB}g^{CD}\partial _{C}\lambda \partial _{D}\lambda ]+d\left( d-1\right)
[\partial _{A}\sigma \partial _{B}\sigma \\ 
\\ 
-\frac{1}{2}g_{AB}g^{CD}\partial _{C}\sigma \partial _{D}\sigma
]+Dd[\partial _{A}\lambda \partial _{B}\sigma +\partial _{A}\sigma \partial
_{B}\lambda \\ 
\\ 
-\frac{1}{2}g_{AB}g^{CD}(\partial _{C}\lambda \partial _{D}\sigma +\partial
_{C}\sigma \partial _{D}\lambda )]-[\bar{R}_{AB}-\frac{1}{2}g_{AB}\bar{R}]
\\ 
\\ 
+\frac{1}{2}g_{AB}[e^{-2\lambda }\tilde{R}+e^{-2\sigma }\hat{R}-2\Lambda ]=0.%
\end{array}
\tag{43c}
\end{equation}

It is known that in two dimensions we can always choose locally the metric as

\begin{equation}
g_{AB}=-N^{2}(x^{C})\delta _{AB}.  \tag{44}
\end{equation}%
Let us assume that Ricci tensor $\bar{R}_{AB}$ vanishes. Then the equation
(43a) is reduced to

\begin{equation}
\begin{array}{c}
D^{2}\left( D-1\right) \partial ^{A}\lambda \partial _{A}\lambda
+Dd(d+1)\partial ^{A}\sigma \partial _{A}\sigma +2Dd\left( D-1\right)
\partial ^{A}\lambda \partial _{A}\sigma \\ 
\\ 
+2D(D-1)\partial ^{A}\partial _{A}\lambda +2Dd\partial ^{A}\partial
_{A}\sigma \\ 
\\ 
-N^{2}[\left( D-2\right) e^{-2\lambda }\tilde{R}+De^{-2\sigma }\hat{R}%
-2D\Lambda ]=0.%
\end{array}
\tag{45a}
\end{equation}%
Similarly, (43b) becomes

\begin{equation}
\begin{array}{c}
d^{2}\left( d-1\right) \partial ^{A}\sigma \partial _{A}\sigma
+Dd(D+1)\partial ^{A}\lambda \partial _{A}\lambda +2Dd\left( d-1\right)
\partial ^{A}\lambda \partial _{A}\sigma \\ 
\\ 
+2d(d-1)\partial ^{A}\partial _{A}\sigma +2Dd\partial ^{A}\partial
_{A}\lambda \\ 
\\ 
-N^{2}[de^{-2\lambda }\tilde{R}+(d-2)e^{-2\sigma }\hat{R}-2d\Lambda ]=0.%
\end{array}
\tag{45b}
\end{equation}%
While (43c) leads

\begin{equation}
\begin{array}{c}
D\left( D-1\right) [\partial _{A}\lambda \partial _{B}\lambda -\frac{1}{2}%
\delta _{AB}\delta ^{CD}\partial _{C}\lambda \partial _{D}\lambda ]+d\left(
d-1\right) [\partial _{A}\sigma \partial _{B}\sigma \\ 
\\ 
-\frac{1}{2}\delta _{AB}\delta ^{CD}\partial _{C}\sigma \partial _{D}\sigma
]+Dd[\partial _{A}\lambda \partial _{B}\sigma +\partial _{A}\sigma \partial
_{B}\lambda \\ 
\\ 
-\frac{1}{2}\delta _{AB}\delta ^{CD}(\partial _{C}\lambda \partial
_{D}\sigma +\partial _{C}\sigma \partial _{D}\lambda )]+\frac{1}{2}%
N^{2}\delta _{AB}[e^{-2\lambda }\tilde{R}+e^{-2\sigma }\hat{R}-2\Lambda ]=0.%
\end{array}
\tag{45c}
\end{equation}%
From (45c) we obtain the equations

\begin{equation}
\begin{array}{c}
\frac{1}{2}D(D-1)[\dot{\lambda}^{2}-(\lambda ^{^{^{\prime }}})^{2}]+\frac{1}{%
2}d(d-1)[\left( \dot{\sigma}\right) ^{2}-(\sigma ^{^{^{\prime }}})^{2}] \\ 
\\ 
+Dd(\dot{\lambda}\dot{\sigma}-\lambda ^{^{\prime }}\sigma ^{^{\prime }})+%
\frac{1}{2}N^{2}(e^{-2\lambda }\tilde{R}+e^{-2\sigma }\hat{R}-2\Lambda )=0,%
\end{array}
\tag{46a}
\end{equation}

\begin{equation}
\begin{array}{c}
\frac{1}{2}D(D-1)[(\lambda ^{^{\prime }})^{2}-(\dot{\lambda})^{2}]+\frac{1}{2%
}d(d-1)[(\sigma ^{^{\prime }})^{2}-(\dot{\sigma})^{2}] \\ 
\\ 
+Dd(\lambda ^{^{\prime }}\sigma ^{^{\prime }}-\dot{\lambda}\dot{\sigma})+%
\frac{1}{2}N^{2}(e^{-2\lambda }\tilde{R}+e^{-2\sigma }\hat{R}-2\Lambda )=0,%
\end{array}
\tag{46b}
\end{equation}%
and

\begin{equation}
\begin{array}{c}
D(D-1)\dot{\lambda}\lambda ^{^{\prime }}+d(d-1)\dot{\sigma}\sigma ^{^{\prime
}}+Dd\{\dot{\lambda}\sigma ^{^{\prime }}+\lambda ^{^{\prime }}\dot{\sigma}%
\}=0.%
\end{array}
\tag{46c}
\end{equation}

For a flat homogeneous internal universe with zero cosmological constant we
have

\begin{equation}
\tilde{R}=0,\hat{R}=0,\Lambda =0.  \tag{47}
\end{equation}%
For this case, (46a) and (46b) lead to the same equation, namely

\begin{equation}
\begin{array}{c}
\frac{1}{2}D(D-1)[\dot{\lambda}^{2}-(\lambda ^{^{^{\prime }}})^{2}]+\frac{1}{%
2}d(d-1)[\left( \dot{\sigma}\right) ^{2}-(\sigma ^{^{^{\prime }}})^{2}] \\ 
\\ 
+Dd(\dot{\lambda}\dot{\sigma}-\lambda ^{^{\prime }}\sigma ^{^{\prime }})=0.%
\end{array}
\tag{48}
\end{equation}%
While (45a) and (45b) become

\begin{equation}
\begin{array}{c}
D\left( D-1\right) \partial ^{A}\lambda \partial _{A}\lambda +d(d+1)\partial
^{A}\sigma \partial _{A}\sigma +2d\left( D-1\right) \partial ^{A}\lambda
\partial _{A}\sigma \\ 
\\ 
+2(D-1)\partial ^{A}\partial _{A}\lambda +2d\partial ^{A}\partial _{A}\sigma
=0%
\end{array}
\tag{49a}
\end{equation}%
and

\begin{equation}
\begin{array}{c}
d\left( d-1\right) \partial ^{A}\sigma \partial _{A}\sigma +D(D+1)\partial
^{A}\lambda \partial _{A}\lambda +2D\left( d-1\right) \partial ^{A}\lambda
\partial _{A}\sigma \\ 
\\ 
+2(d-1)\partial ^{A}\partial _{A}\sigma +2D\partial ^{A}\partial _{A}\lambda
=0,%
\end{array}
\tag{49b}
\end{equation}%
respectively.

Our aim is now to solve the formulae (46c), (48), (49a) and (49b). For that
purpose we write

\begin{equation}
\begin{array}{c}
\lambda (x^{C})=\lambda _{1}(x^{1})+\lambda _{2}(x^{2}) \\ 
\\ 
\sigma (x^{C})=\sigma _{1}(x^{1})+\sigma _{2}(x^{2}).%
\end{array}
\tag{50}
\end{equation}%
We shall look for a solution of the form

\begin{equation}
\begin{array}{c}
\sigma _{1}=\alpha \lambda _{1}, \\ 
\\ 
\sigma _{2}=\beta \lambda _{2},%
\end{array}
\tag{51}
\end{equation}%
where $\alpha $ and $\beta $ are constants. Substituting (50) into (46c) we
find two possibilities depending if

\[
F\equiv \alpha (d-1)+D 
\]%
is zero or different from zero. For $F=0$ the problem is reduced to just one
time. So, in what follows we shall assume that $F\neq 0.$ In this case we
find that $\beta $ is given by

\begin{equation}
\beta =-\frac{[D(D-1)+\alpha Dd]}{\alpha d(d-1)+Dd}.  \tag{52}
\end{equation}

We learn that (51) is consistent with (49a) and (49b) if $\alpha $ and $%
\beta $ satisfy the condition

\begin{equation}
(\alpha -\beta )[Dd(\alpha +\beta )+d(d-1)\alpha \beta +D(D-1)]=0.  \tag{53}
\end{equation}%
The roots of this equation are

\begin{equation}
\beta =\alpha  \tag{54a}
\end{equation}%
and

\begin{equation}
\beta =-\frac{[D(D-1)+\alpha Dd]}{\alpha d(d-1)+Dd}.  \tag{54b}
\end{equation}%
Note that (54b) is consistent with (52). The equation (54a) is obtained from
(54b) when the quantity

\begin{equation}
A=d(d-1)\alpha ^{2}+2Dd\alpha +D(D-1)  \tag{55}
\end{equation}%
vanishes. In fact, one can prove that $\alpha \neq \beta $ implies that $%
A\neq 0.$ In this case, (48) is reduced to%
\begin{equation}
A(\dot{\lambda}_{1})^{2}-B(\lambda _{2}^{^{\prime }})^{2}=0,  \tag{56}
\end{equation}%
where $B$ is defined as

\begin{equation}
B\equiv d(d-1)\beta ^{2}+2Dd\beta +D(D-1).  \tag{57}
\end{equation}%
The solution of (56) is now straightforward. We find%
\begin{equation}
\lambda _{\pm }=C\left[ x^{1}\pm \sqrt{\frac{A}{B}}x^{2}\right] ,  \tag{58}
\end{equation}%
where $C$ is a constant. This expression can also be written as

\begin{equation}
\lambda _{\pm }=C\left[ x^{1}\pm \sqrt{\frac{d[D+(d-1)\alpha ]^{2}}{D(D+d-1)}%
}ix^{2}\right] .  \tag{59}
\end{equation}%
Observe that $\lambda _{\pm }$ is a complex function.

On the other hand for the case $A=0$, from (50) and (51) we find

\begin{equation}
\sigma =\alpha \lambda .  \tag{60}
\end{equation}%
A straightforward computation shows that $\lambda $ can be written in such a
way that

\begin{equation}
\begin{array}{lll}
\ddot{\lambda}_{1}+\frac{D(1-\alpha )}{(d-1)\alpha +D}(\dot{\lambda}_{1})^{2}
& = & \gamma ^{2}, \\ 
&  &  \\ 
\lambda _{2}^{^{\prime \prime }}+\frac{D(1-\alpha )}{(d-1)\alpha +D}(\lambda
_{2}^{^{\prime }})^{2} & = & -\gamma ^{2},%
\end{array}
\tag{61}
\end{equation}%
where $\gamma $ is a constant. Thus, we find that the solutions for the case 
$A=0$ corresponds to two possibilities for the constant $\alpha $, namely

\begin{equation}
\begin{array}{c}
\alpha _{1}=-\frac{D}{d-1}\left[ 1-\sqrt{1-\left( 1-\frac{1}{D}\right)
\left( 1-\frac{1}{d}\right) }\right]%
\end{array}
\tag{62a}
\end{equation}
and

\begin{equation}
\begin{array}{c}
\alpha _{2}=-\frac{D}{d-1}\left[ 1+\sqrt{1-\left( 1-\frac{1}{D}\right)
\left( 1-\frac{1}{d}\right) }\right] .%
\end{array}
\tag{62b}
\end{equation}%
For $\alpha _{1}$ the solution is

\begin{equation}
\begin{array}{c}
\lambda =-\frac{\gamma }{\kappa }t_{1}+\frac{1}{\kappa ^{2}}\ln \left(
Ce^{2\gamma \kappa t_{1}}-1\right) +\frac{1}{\kappa ^{2}}\ln \cos \left(
\gamma \kappa t_{2}-\varphi _{0}\right) .%
\end{array}
\tag{63a}
\end{equation}%
while for $\alpha _{2}$ we have%
\begin{equation}
\begin{array}{c}
\lambda =-\frac{1}{\kappa ^{2}}\ln \cos \left( \gamma \kappa t_{1}-\varphi
_{0}\right) +\frac{\gamma }{\kappa }t_{1}-\frac{1}{\kappa ^{2}}\ln \left(
Ce^{2\gamma \kappa t_{2}}-1\right) .%
\end{array}
\tag{63b}
\end{equation}

Now, if $\gamma =0$ we find

\begin{equation}
\begin{array}{c}
\lambda =\frac{1}{\kappa ^{2}}\ln \left[ \left( t_{1}-\tau _{1}\right)
\left( t_{2}-\tau _{2}\right) \right]%
\end{array}
\tag{64a}
\end{equation}
and

\begin{equation}
\begin{array}{c}
\lambda =\frac{1}{\kappa ^{2}}\ln \left[ \left( t_{1}-\tau _{1}\right)
\left( t_{2}-\tau _{2}\right) \right] .%
\end{array}
\tag{64b}
\end{equation}
Here, the quantities $C,\tau _{1},\tau _{2},\varphi _{0}$ are arbitrary,
while $\kappa $ is defines as

\begin{equation}
\begin{array}{c}
\kappa =\left\vert \frac{D(1-\alpha )}{(d-1)\alpha +D}\right\vert .%
\end{array}
\tag{65}
\end{equation}

It is worth mentioning that for the description of our universe the solution
(59) is of special interest because it leads to a universe with open first
time $x_{1}$ and compact second time $x_{2}$ (see [7]).

\bigskip\ 

\noindent \textbf{6.- FINAL REMARKS}

\smallskip\ 

In this article we have studied duality symmetries in Kaluza-Klein $n+D+d$
dimensional cosmological models\textbf{. }We first briefly reviewed the case 
$1+D+d$\textbf{\ }cosmological model. We wrote the action of this model in
such a way that the duality symmetry becomes manifest. As a particular case,
we discussed both at the classical and the quantum level the $1+3+1$
cosmological model. In section 4, we studied, from the point of view of
duality, the more general case of a $n+D+d$ cosmological model. We
discovered that, except for the case $n=2,$ the Einstein-Hilbert action in $%
n+D+d$ dimensions is invariant under the duality symmetry $a\rightarrow 
\frac{1}{a}$ and $b\rightarrow \frac{1}{b}.$ We studied the $2+D+d$
cosmological model in some detail finding an explicit classical solution.
One of the interesting features of the $2+D+d$ cosmological model is that,
in spite of lacking a duality symmetry, it leads to a universe in which the
second time can be considered as a compact time-like dimension, while the
first usual time behaves as an open dimension. It turns out that this kind
of solution was already anticipated by Bars and Kounnas [7].

It is clear from the present results that the traditional
Friedmann-Robertson-Walker cosmological model is contained in the $2+D+d$
cosmological model. The question arises whether other traditional
cosmological models such as the different Bianchi models are also contained
in the $2+D+d$ model. The really interesting problem, however, is to find a
mechanism to decide whether the $2+D+d$ model is the correct model of the
universe. Experimentally, it is an interesting possibility because
presumably the second time is shrinking to zero in the first stage of the
evolution of the universe, leading after that to the usual evolving
universe. Theoretically, one becomes intriguing why duality is broken in the
case of two times cosmological model, distinguishing the $2+D+d$ model of
other $n+D+d$ models. In this work we tried to understand classically this
interesting feature of the $2+D+d$ cosmological model but beyond of finding
a consistent solution with the present evolution of our universe there seems
not to be a clear reason why the duality symmetry is broken.

An open problem for further research is to quantize the $n+D+d$ cosmological
model. In this case it may be interesting to see what are the consequences
of the duality symmetry in the corresponding Wheeler-de Witt equation and in
the associated state of the universe.

\newpage

\textbf{APPENDIX A}

\bigskip\ 

From (1) we have that the only nonvanishing elements of the $1+D+d$
dimensional metric $g_{\alpha \beta }$, with $\alpha ,\beta =0,1...,1+D+d$
are

\begin{equation}
\begin{array}{lll}
g_{00} & = & -N^{2}, \\ 
&  &  \\ 
g_{ij} & = & a^{2}(t)\tilde{g}_{ij}, \\ 
&  &  \\ 
g_{ab} & = & b^{2}(t)\hat{g}_{ab},%
\end{array}
\tag{A1}
\end{equation}%
where the metric $\tilde{g}_{ij}$ corresponds to the $D$-dimensional
homogenous space, while $\hat{g}_{ab}$ is metric of the $d$-dimensional
homogeneous space$.$

We find that the only non-vanishing Christoffel symbols

\begin{equation}
\Gamma _{\alpha \beta }^{\mu }=\frac{1}{2}g^{\mu \nu }(g_{\nu \alpha ,\beta
}+g_{\nu \beta ,\alpha }-g_{\alpha \beta ,\nu })  \tag{A2}
\end{equation}%
are

\begin{equation}
\begin{array}{lll}
\Gamma _{ij}^{0} & = & N^{-2}a\dot{a}\tilde{g}_{ij}, \\ 
&  &  \\ 
\Gamma _{j0}^{i} & = & a^{-1}\dot{a}\delta _{j}^{i}, \\ 
&  &  \\ 
\Gamma _{00}^{0} & = & N^{-1}\dot{N}, \\ 
&  &  \\ 
\Gamma _{jk}^{i} & = & \tilde{\Gamma}_{jk}^{i}, \\ 
&  &  \\ 
\Gamma _{ab}^{0} & = & N^{-2}b\dot{b}\hat{g}_{ab}, \\ 
&  &  \\ 
\Gamma _{b0}^{a} & = & b^{-1}\dot{b}\delta _{b}^{a}, \\ 
&  &  \\ 
\Gamma _{bc}^{a} & = & \hat{\Gamma}_{bc}^{a}.%
\end{array}
\tag{A3}
\end{equation}%
Using (A3) we discover that the only non-vanishing components of the Riemann
tensor

\begin{equation}
R_{\nu \alpha \beta }^{\mu }=\Gamma _{\nu \beta ,\alpha }^{\mu }-\Gamma
_{\nu \alpha ,\beta }^{\mu }+\Gamma _{\sigma \alpha }^{\mu }\Gamma _{\nu
\beta }^{\sigma }-\Gamma _{\sigma \beta }^{\mu }\Gamma _{\nu \alpha
}^{\sigma }  \tag{A4}
\end{equation}%
are

\begin{equation}
\begin{array}{lll}
R_{i0j}^{0} & = & (N^{-2}a\ddot{a}-N^{-3}\dot{N}a\dot{a})\tilde{g}_{ij}, \\ 
&  &  \\ 
R_{0j0}^{i} & = & (-a^{-1}\ddot{a}+N^{-1}\dot{N}a^{-1}\dot{a})\delta
_{j}^{i}, \\ 
&  &  \\ 
R_{jkl}^{i} & = & \tilde{R}_{jkl}^{i}+N^{-2}\dot{a}^{2}(\delta _{k}^{i}%
\tilde{g}_{jl}-\delta _{l}^{i}\tilde{g}_{jk}), \\ 
&  &  \\ 
R_{a0b}^{0} & = & (N^{-2}b\ddot{b}-N^{-3}\dot{N}b\dot{b})\hat{g}_{ab}, \\ 
&  &  \\ 
R_{0b0}^{a} & = & (-b^{-1}\ddot{b}+N^{-1}\dot{N}b^{-1}\dot{b})\delta
_{b}^{a}, \\ 
&  &  \\ 
R_{bcd}^{a} & = & \hat{R}_{bcd}^{a}+N^{-2}\dot{b}^{2}(\delta _{c}^{a}\hat{g}%
_{bd}-\delta _{d}^{a}\hat{g}_{bc}), \\ 
&  &  \\ 
R_{ajb}^{i} & = & N^{-2}a^{-1}\dot{a}b\dot{b}\delta _{j}^{i}\hat{g}_{ab}, \\ 
&  &  \\ 
R_{ibj}^{a} & = & N^{-2}a\dot{a}b^{-1}\dot{b}\delta _{b}^{a}\tilde{g}_{ij}.%
\end{array}
\tag{A5}
\end{equation}

From (A5) we get the non-vanishing components of the Ricci tensor $R_{\mu
\nu }=R_{\mu \alpha \nu }^{\alpha }$;

\begin{equation}
\begin{array}{lll}
R_{00} & = & -Da^{-1}\ddot{a}+DN^{-1}\dot{N}a^{-1}\dot{a}-db^{-1}\ddot{b}%
+dN^{-1}\dot{N}b^{-1}\dot{b}, \\ 
&  &  \\ 
R_{ij} & = & (N^{-2}a\ddot{a}-N^{-3}\dot{N}a\dot{a}+(D-1)N^{-2}\dot{a}%
^{2}+dN^{-2}a\dot{a}b^{-1}\dot{b})\tilde{g}_{ij}+\tilde{R}_{ij}, \\ 
&  &  \\ 
R_{ab} & = & (N^{-2}b\ddot{b}-N^{-3}\dot{N}b\dot{b}+(d-1)N^{-2}\dot{b}%
^{2}+DN^{-2}a^{-1}\dot{a}b\dot{b})\hat{g}_{ab}+\hat{R}_{ab}.%
\end{array}
\tag{A6}
\end{equation}%
Thus, the Ricci scalar $R=g^{\mu \nu }R_{\mu \nu }$ is given by

\begin{equation}
\begin{array}{lll}
R & = & 2DN^{-2}a^{-1}\ddot{a}-2DN^{-3}\dot{N}a^{-1}\dot{a}%
+D(D-1)N^{-2}a^{-2}\dot{a}^{2}+D(D-1)k_{1}a^{-2} \\ 
&  &  \\ 
&  & +2dN^{-2}b^{-1}\ddot{b}-2dN^{-3}\dot{N}b^{-1}\dot{b}+d(d-1)N^{-2}b^{-2}%
\dot{b}^{2}+d(d-1)k_{2}b^{-2} \\ 
&  &  \\ 
&  & +2dDN^{-2}a^{-1}\dot{a}b^{-1}\dot{b}.%
\end{array}
\tag{A7}
\end{equation}

\bigskip\ 

\bigskip\ 

\bigskip\ 

\noindent \textbf{APPENDIX B}

\bigskip\ 

Assuming $1+3+1$ dimensions we observe that (A1) is reduced to%
\begin{equation}
\begin{array}{lll}
g_{00} & = & -N^{2}, \\ 
&  &  \\ 
g_{ij} & = & a^{2}(t)\tilde{g}_{ij}, \\ 
&  &  \\ 
g_{44} & = & b^{2}(t),%
\end{array}
\tag{B1}
\end{equation}%
where the metric $\tilde{g}_{ij}$ corresponds to the spatial sector, with $%
i,j=1,2,3.$

Considering (B1) we find that the only non-vanishing Christoffel symbols

\begin{equation}
\Gamma _{\alpha \beta }^{\mu }=\frac{1}{2}g^{\mu \nu }(g_{\nu \alpha ,\beta
}+g_{\nu \beta ,\alpha }-g_{\alpha \beta ,\nu })  \tag{B2}
\end{equation}
are

\begin{equation}
\begin{array}{lll}
\Gamma _{ij}^{0} & = & N^{-2}a\dot{a}\tilde{g}_{ij}, \\ 
&  &  \\ 
\Gamma _{j0}^{i} & = & a^{-1}\dot{a}\delta _{j}^{i}, \\ 
&  &  \\ 
\Gamma _{00}^{0} & = & N^{-1}\dot{N}, \\ 
&  &  \\ 
\Gamma _{jk}^{i} & = & \tilde{\Gamma}_{jk}^{i}, \\ 
&  &  \\ 
\Gamma _{44}^{0} & = & N^{-2}b\dot{b}, \\ 
&  &  \\ 
\Gamma _{04}^{4} & = & b^{-1}\dot{b}.%
\end{array}
\tag{B3}
\end{equation}%
Using (B3) we discover that the only non-vanishing components of the Riemann
tensor

\begin{equation}
R_{\nu \alpha \beta }^{\mu }=\Gamma _{\nu \beta ,\alpha }^{\mu }-\Gamma
_{\nu \alpha ,\beta }^{\mu }+\Gamma _{\sigma \alpha }^{\mu }\Gamma _{\nu
\beta }^{\sigma }-\Gamma _{\sigma \beta }^{\mu }\Gamma _{\nu \alpha
}^{\sigma }  \tag{B4}
\end{equation}%
are

\begin{equation}
\begin{array}{lll}
R_{i0j}^{0} & = & (N^{-2}a\ddot{a}-N^{-3}\dot{N}a\dot{a})\tilde{g}_{ij}, \\ 
&  &  \\ 
R_{0j0}^{i} & = & (-a^{-1}\ddot{a}+N^{-1}\dot{N}a^{-1}\dot{a})\delta
_{j}^{i}, \\ 
&  &  \\ 
R_{jkl}^{i} & = & \tilde{R}_{jkl}^{i}+N^{-2}\dot{a}^{2}(\delta _{k}^{i}%
\tilde{g}_{jl}+\delta _{l}^{i}\tilde{g}_{jk}), \\ 
&  &  \\ 
R_{040}^{4} & = & -b^{-1}\ddot{b}+N^{-1}\dot{N}b^{-1}\dot{b}, \\ 
&  &  \\ 
R_{404}^{0} & = & N^{-2}b\ddot{b}-N^{-3}\dot{N}b\dot{b}, \\ 
&  &  \\ 
R_{4j4}^{i} & = & N^{-2}a^{-1}\dot{a}b\dot{b}\delta _{j}^{i}, \\ 
&  &  \\ 
R_{i4j}^{4} & = & N^{-2}a\dot{a}b^{-1}\dot{b}\tilde{g}_{ij}.%
\end{array}
\tag{B5}
\end{equation}

From (B5) we get the non-vanishing components of the Ricci tensor $R_{\mu
\nu }=R_{\mu \alpha \nu }^{\alpha }$;

\begin{equation}
\begin{array}{lll}
R_{00} & = & -3a^{-1}\ddot{a}+3N^{-1}\dot{N}a^{-1}\dot{a}-b^{-1}\ddot{b}%
+N^{-1}\dot{N}b^{-1}\dot{b}, \\ 
&  &  \\ 
R_{ij} & = & (N^{-2}a\ddot{a}-N^{-3}\dot{N}a\dot{a}+2N^{-2}\dot{a}%
^{2}+N^{-2}a\dot{a}b^{-1}\dot{b})\tilde{g}_{ij}+\tilde{R}_{ij}, \\ 
&  &  \\ 
R_{44} & = & N^{-2}b\ddot{b}-N^{-3}\dot{N}b\dot{b}+3N^{-2}a^{-1}\dot{a}b\dot{%
b}.%
\end{array}
\tag{B6}
\end{equation}%
Thus, the Ricci scalar $R=g^{\mu \nu }R_{\mu \nu }$ is given by

\begin{equation}
\begin{array}{lll}
R & = & 6N^{-2}a^{-1}\ddot{a}-6N^{-3}\dot{N}a^{-1}\dot{a}+6N^{-2}a^{-2}\dot{a%
}^{2}+6ka^{-2}+2N^{-2}b^{-1}\ddot{b} \\ 
&  &  \\ 
&  & -2N^{-3}\dot{N}b^{-1}\dot{b}+6N^{-2}\dot{N}a^{-1}\dot{a}b^{-1}\dot{b}.%
\end{array}
\tag{B7}
\end{equation}

\bigskip

\noindent \textbf{APPENDIX C}

\bigskip\ 

Consider the $n+D+d$ dimensional metric $g_{\alpha \beta }$, with $\alpha
,\beta =0,1...,n,n+1,...,n+D+d$%
\begin{equation}
\begin{array}{lll}
g_{AB} & = & \bar{g}_{AB}(x^{C}), \\ 
&  &  \\ 
g_{ij} & = & a^{2}(x^{C})\tilde{g}_{ij}, \\ 
&  &  \\ 
g_{ab} & = & b^{2}(x^{C})\hat{g}_{ab},%
\end{array}
\tag{C1}
\end{equation}%
where the metric $\tilde{g}_{ij}$ corresponds to the $D$-dimensional
homogenous space, while $\hat{g}_{ab}$ is metric of the $d$-dimensional
homogeneous space$.$ Furthermore, the indices $A,B...etc$ run from $1$ to $%
n, $ the indices $i,j...etc$ run from $n+1$ to $n+D$ and the indices $%
a,b...etc$ run from $n+D+1$ to $n+D+d.$

We find that the only non-vanishing Christoffel symbols

\begin{equation}
\Gamma _{\alpha \beta }^{\mu }=\frac{1}{2}g^{\mu \nu }(g_{\nu \alpha ,\beta
}+g_{\nu \beta ,\alpha }-g_{\alpha \beta ,\nu })  \tag{C2}
\end{equation}
are

\begin{equation}
\begin{array}{lll}
\Gamma _{ij}^{A} & = & -g^{AB}a\partial _{B}a\tilde{g}_{ij}, \\ 
&  &  \\ 
\Gamma _{jA}^{i} & = & a^{-1}\partial _{A}a\delta _{j}^{i}, \\ 
&  &  \\ 
\Gamma _{BC}^{A} & = & \bar{\Gamma}_{BC}^{A}, \\ 
&  &  \\ 
\Gamma _{jk}^{i} & = & \tilde{\Gamma}_{jk}^{i}, \\ 
&  &  \\ 
\Gamma _{ab}^{A} & = & -g^{AB}b\partial _{B}b\hat{g}_{ab}, \\ 
&  &  \\ 
\Gamma _{bA}^{a} & = & b^{-1}\partial _{A}b\delta _{b}^{a}, \\ 
&  &  \\ 
\Gamma _{bc}^{a} & = & \hat{\Gamma}_{bc}^{a}.%
\end{array}
\tag{C3}
\end{equation}%
Using (C3) we discover that the only non-vanishing components of the Riemann
tensor

\begin{equation}
R_{\nu \alpha \beta }^{\mu }=\Gamma _{\nu \beta ,\alpha }^{\mu }-\Gamma
_{\nu \alpha ,\beta }^{\mu }+\Gamma _{\sigma \alpha }^{\mu }\Gamma _{\nu
\beta }^{\sigma }-\Gamma _{\sigma \beta }^{\mu }\Gamma _{\nu \alpha
}^{\sigma }  \tag{C4}
\end{equation}%
are

\begin{equation}
\begin{array}{lll}
R_{BCD}^{A} & = & \bar{R}_{BCD}^{A} \\ 
&  &  \\ 
R_{iBj}^{A} & = & (-aD_{B}\partial ^{A}a)\tilde{g}_{ij}, \\ 
&  &  \\ 
R_{AjB}^{i} & = & (-a^{-1}D_{B}\partial _{A}a)\delta _{j}^{i}, \\ 
&  &  \\ 
R_{jkl}^{i} & = & \tilde{R}_{jkl}^{i}-g^{AB}\partial _{A}a\partial
_{B}a(\delta _{k}^{i}\tilde{g}_{jl}-\delta _{l}^{i}\tilde{g}_{jk}), \\ 
&  &  \\ 
R_{aBb}^{A} & = & (-bD_{B}\partial ^{A}b)\hat{g}_{ab}, \\ 
&  &  \\ 
R_{AbB}^{a} & = & (-b^{-1}D_{B}\partial _{A}b)\delta _{b}^{a}, \\ 
&  &  \\ 
R_{bcd}^{a} & = & \hat{R}_{bcd}^{a}-g^{AB}\partial _{A}b\partial
_{B}b(\delta _{c}^{a}\hat{g}_{bd}-\delta _{d}^{a}\hat{g}_{bc}), \\ 
&  &  \\ 
R_{ajb}^{i} & = & -(a^{-1}b)g^{AB}\partial _{A}a\partial _{B}b\delta _{j}^{i}%
\hat{g}_{ab}, \\ 
&  &  \\ 
R_{ibj}^{a} & = & -(b^{-1}a)g^{AB}\partial _{A}a\partial _{B}b\delta _{b}^{a}%
\tilde{g}_{ij}.%
\end{array}
\tag{C5}
\end{equation}

From (C5) we get the non-vanishing components of the Ricci tensor $R_{\mu
\nu }=R_{\mu \alpha \nu }^{\alpha }$;

\begin{equation}
\begin{array}{lll}
R_{AB} & = & \bar{R}_{AB}-Da^{-1}D_{B}\partial _{A}a-db^{-1}D_{B}\partial
_{A}b, \\ 
&  &  \\ 
R_{ij} & = & -(aD_{A}\partial ^{A}a+(D-1)g^{AB}\partial _{A}a\partial
_{B}a+d(b^{-1}a)g^{AB}\partial _{A}a\partial _{B}b)\tilde{g}_{ij}+\tilde{R}%
_{ij}, \\ 
&  &  \\ 
R_{ab} & = & -(bD_{A}\partial ^{A}b+(d-1)g^{AB}\partial _{A}b\partial
_{B}b+D(a^{-1}b)g^{AB}\partial _{A}a\partial _{B}b)\hat{g}_{ab}+\hat{R}_{ab}.%
\end{array}
\tag{C6}
\end{equation}%
Thus, the Ricci scalar $R=g^{\mu \nu }R_{\mu \nu }$ is given by

\begin{equation}
\begin{array}{lll}
R & = & -2Da^{-1}D_{A}\partial ^{A}a-D(D-1)g^{AB}a^{-2}\partial
_{A}a\partial _{B}a \\ 
&  &  \\ 
&  & -2db^{-1}D_{A}\partial ^{A}b-d(d-1)g^{AB}b^{-2}\partial _{A}b\partial
_{B}b \\ 
&  &  \\ 
&  & -2Dd(b^{-1}a^{-1})g^{AB}\partial _{A}a\partial _{B}b+\bar{R}+a^{-2}%
\tilde{R}+b^{-2}\hat{R}.%
\end{array}
\tag{C7}
\end{equation}

\end{document}